\documentclass[aip,apl,numerical,reprint]{revtex4-1}
\usepackage{graphicx}%
\usepackage{bm}%
\usepackage{amssymb}
\usepackage[amssymb]{SIunits}
\usepackage[version=3]{mhchem}
\usepackage{color}
\usepackage{textcomp}

\newcommand{\CNRSAddress}{Institut N\'eel, CNRS and Universit\'e de Grenoble, BP 166, 25 rue des Martyrs, 38042 Grenoble, France}
\newcommand{\CEAddress}{INAC, CEA and Universit\'e de Grenoble, 17 rue des Martyrs, 38054 Grenoble, France}

\begin{document}

\title{Optical properties of single ZnTe nanowires grown at low temperature}

\author{A. Artioli}
\affiliation{\CNRSAddress}
\author{P. Rueda-Fonseca}
\email{pamela.rueda@grenoble.cnrs.fr}
\affiliation{\CNRSAddress}
\affiliation{\CEAddress}
\author{P. Stepanov}
\affiliation{\CNRSAddress}
\author{E. Bellet-Amalric}
\affiliation{\CEAddress}
\author{M. Den Hertog}
\affiliation{\CNRSAddress}
\author{C. Bougerol}
\affiliation{\CNRSAddress}
\author{Y. Genuist}
\affiliation{\CNRSAddress}
\author{F. Donatini}
\affiliation{\CNRSAddress}
\author{R. Andr\'e}
\affiliation{\CNRSAddress}
\author{G. Nogues}
\affiliation{\CNRSAddress}
\author{K. Kheng}
\affiliation{\CEAddress}
\author{S. Tatarenko}
\affiliation{\CNRSAddress}
\author{D. Ferrand}
\affiliation{\CNRSAddress}
\author{J. Cibert}
\affiliation{\CNRSAddress}
\date{\today{}}



\begin{abstract}
Optically active gold-catalyzed ZnTe nanowires have been grown by
molecular beam epitaxy, on a ZnTe(111) buffer layer, at low
temperature (\unit{350}{\degree C}) under Te rich conditions, and at
ultra-low density (from 1 to 5 nanowires per \unit{}{\micro \square
\meter}). The crystalline structure is zinc blende as identified by
transmission electron microscopy. All nanowires are tapered and the
majority of them are $\langle 111\rangle$ oriented. Low temperature
micro-photoluminescence and cathodoluminescence experiments have
been performed on single nanowires. We observe a narrow emission
line with a blue-shift of 2 or 3 meV with respect to the  exciton
energy in bulk ZnTe. This shift is attributed to the strain induced
by a 5~nm-thick oxide layer covering the nanowires, and this
assumption is supported by a quantitative estimation of the strain
in the nanowires.
\end{abstract}

\maketitle

There is currently a wide-spread interest for semiconductor
nanowires (NWs), driven by their potential to constitute suitable
building blocks for future nanoelectronic and nanophotonic devices.
\cite{Duan01} During the past decade, selenide and telluride II-VI
NWs have been extensively investigated for various applications such
as nano-pillar solar cells,\cite{Fan09} photodetectors \cite{Meng09}
or single photon sources.\cite{Boun12} Among II-VI's, ZnTe based NWs
are particularly promising as offering a large range of
potentialities. They can be grown by molecular beam epitaxy (MBE)
using gold particles as catalyst. They can be efficiently doped
electrically. \cite{Liu13} They can also be doped with magnetic
impurities, \cite{Rada05,Wojn12} while the large difference between
the temperatures suitable for the growth of GaAs NWs and for the
incorporation of Mn as substitutional impurities makes the growth of
(Ga,Mn)As NWs challenging.\cite{Bouravleuv13} As (Zn,Mn)Te can be
doped strongly p-type so that ferromagnetism appears and transport
studies are feasible in 2D,\cite{Ferrand01} ZnTe-based NWs are
attractive for a basic study of spintronics mechanisms in 1D. In
addition, CdTe quantum dots can be incorporated and used as a single
photon source \cite{Wojn11} or as a very sensitive optical probe of
the spin properties. \cite{Beso04}

A good control and the optimization of the growth conditions are
prerequisites to improve the electronic and optical properties of
the NWs. We present the MBE growth and the structural analysis of
ultra-low density ZnTe NWs and we show that their high crystalline
quality allows us to observe micro-photoluminescence ($\mu$PL) and
cathodoluminescence (CL) emission from single NWs.

ZnTe NWs were grown by MBE, using gold particles as a catalyst. A
thin layer of gold was deposited on a 500 nm-thick ZnTe(111) buffer
layer previously grown on a GaAs(111)B substrate. Gold droplets were
formed at \unit{350}{\degree C} and the NWs were grown at the same
temperature. As we observed that Te-rich conditions result in a
larger diffusion length on the (111) ZnTe surface,\cite{Rueda13} the
present NWs were grown with a Zn:Te flux ratio 1:2.3, for 30 or 60
min. More details are given in Ref.~\onlinecite{Rueda13}.

\begin{figure}
 \centering
 \includegraphics[width=0.95 \columnwidth] {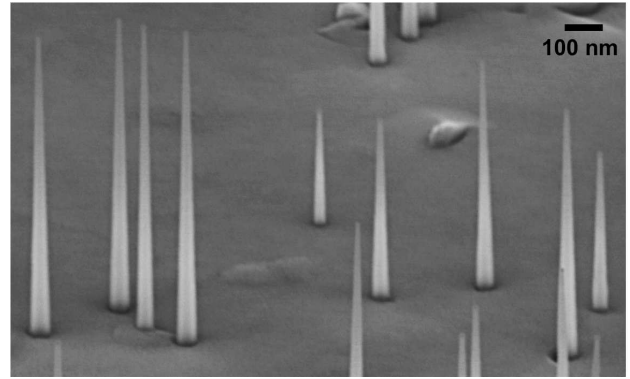}
\caption{SEM image of ZnTe NWs grown at \unit{350}{\degree C} for 30
min under Te rich conditions.} \label{ZnTeNWs}
\end{figure}

Typical ZnTe NWs have been imaged by scanning electron microscopy
(SEM), see Fig.~\ref{ZnTeNWs}. This sample presents an
ultra-low-density of NWs, down to 2 NWs per \unit{}{\micro \square
\meter}. These NWs grown under Te-rich conditions are tapered, with
a thick base (50-70 nm) and a  thin tip (15 nm roughly equal to the
gold particle diameter) . This cone shape is due to lateral growth
induced by the low growth temperature.\cite{Rueda13}  The NW height
distribution is large, from \unit{300}{\nano\meter} to
\unit{1.5}{\micro\meter} for a 30 min growth time.

\begin{figure}
 \centering
 \includegraphics[width=\columnwidth]{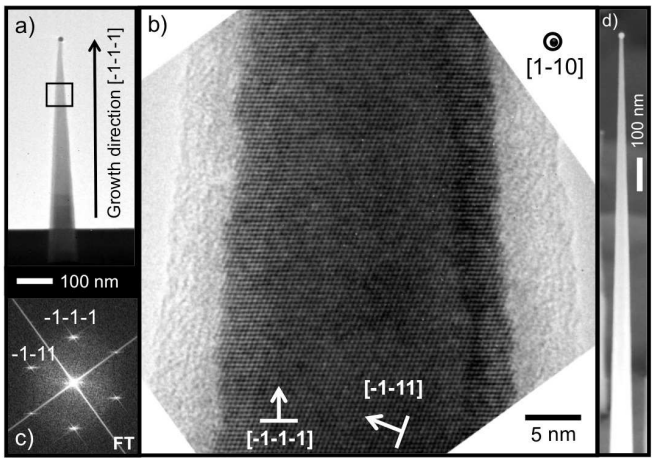}
\caption{(a) Low Resolution TEM image of a ZnTe NW grown for 30 min.
(b) High Resolution TEM image, and (c) Fourier transform. (d) A ZnTe
NW grown for 60 min, at the same scale than in (a).} \label{TEM}
\end{figure}

The majority (about two thirds) of the NWs are oriented $\langle 111
\rangle$, and most of the remaining NWs oriented $\langle 112
\rangle$. Transmission Electron Microscopy (TEM) was performed on a
Philips CM300 microscope equipped with a CCD camera and operated at
300~kV. Images reveal a zinc-blende crystal structure, see
Fig.~\ref{TEM}. Some NWs do not present any defect, but many of them
show twins, as frequently observed in $\langle 111 \rangle$
zinc-blende NWs. We systematically observe an amorphous layer (5 nm
thick) all around the NW, which we attribute to oxidation after
growth (see Fig.~\ref{TEM}b). ZnTe(111) surfaces tend to be rapidly
oxidized either by forming \ce{Te O_2} or/and ZnO
\cite{Ebina1978_ZnTeOxide}. In the case of ZnTe NWs, the formation
of a partially crystallized ZnO shell has been
reported,\cite{Kirmse08} although the presence of Te-oxide in the
outermost part was also suggested by electron energy loss
spectroscopy analysis. In our samples, energy dispersive x-ray
analysis (EDX) also detects the presence of Zn oxide and of some Te
oxide. \cite{Rueda13}

The cone shape of these NWs indicates that, under these growth
conditions (low temperature and Te excess), the diffusion length of
adatoms on the facets of the NW is not large. \cite{Rueda13} This
could limit the range of NW length which can be achieved.
Nevertheless, NWs grown for a longer time (60 min) exhibit a larger
length, as shown in Fig.~\ref{TEM}.d. For the same size of the gold
droplets, the average diameter at the base of the NWs was increased
to \unit{80-100} {\nano\meter}, and the length distribution to
0.8-\unit{2.1}{\micro\meter}.

\begin{figure}
 \centering
 \includegraphics[width= \columnwidth]{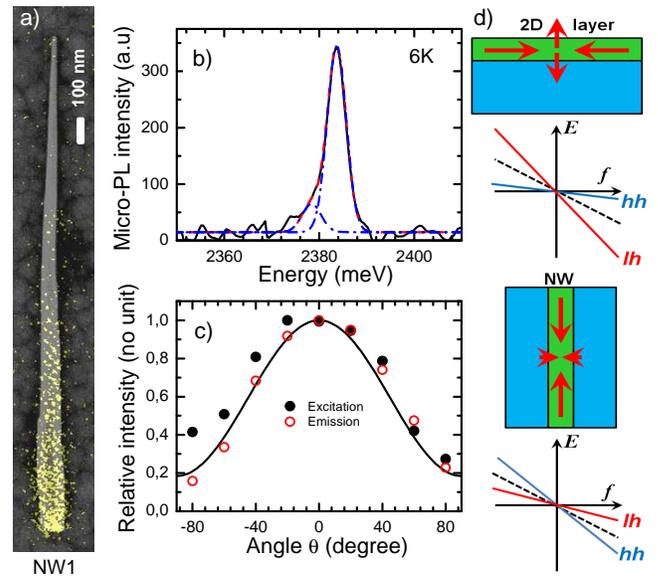}
\caption{(a) Superposition of the high resolution SEM image and the
CL image (yellow pixels) of a ZnTe NW deposited on a patterned
silicon substrate. (b) \textmu PL spectra of the NW shown in (a),
recorded at 6K using a 488 nm laser excitation at $25  \mu $W/$\mu
$m$^2$. The dashed line is a Gaussian fit with a main peak at 2383.7
meV and a satellite line at 2378.4 meV. (c) Dependence of the PL
intensity on the angle $\theta$, shown in (a), between the NW axis
and the direction of the linear polarization of the laser excitation
(closed circles) or the emitted light (open circles). The solid line
is the dependence expected from dielectric screening. (d) Strain
imposed by the substrate on an epitaxial layer (top) and by the
shell on the NW core (bottom), schematically depicted by the arrows,
and their effect on the heavy hole (hh) and light hole (lh)
recombination energy (and the average, dashed line) as a function of
lattice mismatch.}\label{fig:fig3}
\end{figure}

The optical properties of single NWs have been characterized by low
temperature \textmu PL and CL. ZnTe NWs were first deposited on a
patterned silicon substrate by rubbing as-grown samples on the
silicon surface. Isolated NWs were identified and precisely located
on the substrate using a high-resolution SEM, see
Fig.~\ref{fig:fig3}.a. Then low temperature (6~K) spectra of these
NWS were recorded using a cold-finger cryostat and a confocal
\textmu PL set-up. The NWs were excited by \unit{100}{\micro}W of a
\unit{488}{\nano\meter} cw laser beam focused to
\unit{4}{\micro\square\meter} on the selected single NW using a
microscope objective and a piezoelectric scanner. The light emitted
by the NW was collected by the microscope objective and sent to a
\unit{0.46}{\meter} Jobin-Yvon spectrometer equipped with a CCD
camera.

CL images of the same NWs deposited on the silicon substrate were
recorded using an FEI Quanta 200 SEM equipped with a CL accessory
and a low temperature Gatan stage able to cool down the sample to
6~K. \cite{Dona10} The single NW was excited by an electron beam of
\unit{30}{\kilo\electronvolt} and a current of typically
\unit{250}{\pico\ampere}. In comparison with high resolution
field-effect SEM images, the spatial resolution of the CL images is
limited by the low spatial resolution of the thermionic SEM. The CL
light was collected by a parabolic mirror and sent to an avalanche
photodiode synchronized with the electron beam scan.


As shown in Fig.~\ref{fig:fig3}.b, \textmu PL spectra are dominated
by a single emission peak close to the ZnTe exciton band edge. The
spectrum has been fitted using two Gaussian lines having a full
width at half maximum equal to \unit{4}{\milli \electronvolt}. The
main peak is at \unit{2383.7}{\milli \electronvolt}, very close to
the bulk exciton at \unit{2381}{\milli \electronvolt}:
\cite{LeSi89,Cama02} we will discuss it below as originating from a
free or slightly localized exciton. The small satellite peak,
observed \unit{5}{\milli \electronvolt} below the main peak is
attributed to bound excitons. \cite{Cama02}

These results apparently contrast with previous studies performed
with ZnTe/(Zn,Mg)Te core-shell NWs, \cite{Wojn12} where a band edge
emission is observed at \unit{2.31}{\electronvolt}. A line at lower
energy might be ascribed to trapped excitons or complexes, but the
position of the line in the present study, slightly above the
exciton in bulk ZnTe, cannot be ascribed to the influence of lateral
confinement, which remains negligible for the present range of NW
diameters. Indeed such a difference in the exciton energy is
expected from the high sensitivity of the band edge emission to the
strain induced by the presence of a shell around the NW.

According to Ref.~\onlinecite{Aifa07}, the stress induced in the
core by a lattice mismatched shell covering an infinitely long
cylindrical NW is uniform, equal to: $\sigma_\parallel=2 \times
\sigma_\perp=\frac{18 \mu K}{3K+4\mu} (1-\frac{D_c^2}{D_s^2})f.$
Here $\sigma_\parallel$ is the component of the core stress parallel
to the NW, and $\sigma_\perp$ the perpendicular component; $K$ is
the bulk modulus and $\mu$ is the shear modulus of the core and the
shell materials, assumed to have the same isotropic elastic
properties; $f$ is the lattice mismatch between the core and the
shell ($f>0$ if the shell has a larger lattice parameter); $D_c$ is
the core diameter and $D_s$ the external diameter of the core-shell
structure. For semiconductor NWs, a more complete
calculation,\cite{Cibe13} which is beyond the scope of this paper,
shows that the previous expressions can be used for
$\langle111\rangle$ oriented semiconductor NWs using the bulk
modulus $K=(c_{11}+2 c_{12})/3$ and a shear modulus $\bar
{\mu}=\frac{1}{4} \frac{c_{11}-c_{12}}{2}+\frac{3}{4} c_{44}$, where
$c_{11}$, $c_{12}$ and $c_{44}$ are the stiffness coefficients of a
cubic semiconductor. This leads to an hydrostatic shift of the band
edge emission $\Delta_{hyd}$ and a heavy hole-light hole splitting
$\Delta_{HL}$ proportional to the lattice mismatch $f$, reduced by
the geometrical factor $(1-\frac{D_c^2}{D_s^2})$, \emph{i.e}., the
ratio of the shell to the total cross section areas :
$\Delta_{hyd}=-a \frac{12 \bar {\mu}}{3K+4\bar
{\mu}}(1-\frac{D_c^2}{D_s^2})f$, and
$\Delta_{HL}=-\frac{d}{\sqrt{3}} \frac{9 K}{3K+4\bar
{\mu}}(1-\frac{D_c^2}{D_s^2})f$.

For the deformation potentials in ZnTe, we take the values
\cite{LeSi89} $a$=5.3~eV and $d/\sqrt3$=2.5~eV; and for the
stiffness coefficients, \cite{Berlincourt} $c_{11}$=73.7~GPa,
$c_{12}$=42.3~GPa, and $c_{44}$=32.1~GPa. Then the excitonic
emission of a $\langle 111 \rangle$ oriented cubic ZnTe NW is (in
meV, with $f$ in \%) $E_{NW}=2381-88 (1-\frac{D_c^2}{D_s^2})f$ for
the heavy hole and $E_{NW}=2381-44 (1-\frac{D_c^2}{D_s^2})f$ for the
light hole. The heavy hole is the ground state if $f>0$,
\emph{i.e.}, as noticed in Ref.~\onlinecite{Wojn12}, if the core is
under tensile strain.

\begin{figure}
 \centering
 \includegraphics[width=\columnwidth]{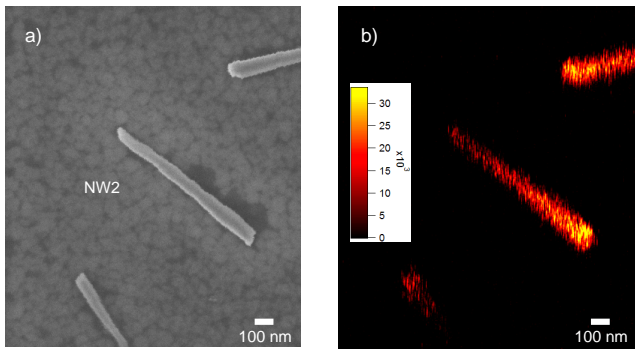}
\caption{(a) High resolution SEM image of 3 broken ZnTe NWs
deposited on a silicon substrate. (b) CL image of the NWs shown in
(a) recorded at 6~K.}\label{fig:fig4}
\end{figure}

For the NWs studied in Ref.~\onlinecite{Wojn12}, with $D_c=$
\unit{70}{\nano \meter}, $D_s=$\unit{130}{\nano \meter}, and
$f=1.04\%$ corresponding to the lattice mismatch between a ZnTe core
and a Zn$_{0.8}$Mg$_{0.2}$Te shell,\cite{Hartmann} we obtain
\unit{2.31}{\electronvolt} for the heavy-hole exciton, in agreement
with the observed PL line. One can note that in this core-shell
geometry, the position of the heavy-hole emission is more sensitive
to the strain than in an epitaxial layer under compressive biaxial
strain: the shifts of the heavy-hole state due to the hydrostatic
and the shear strain add, while they partially compensate in the
epitaxial layer. For the NWs studied in this work, we expect a
compressive strain induced by the oxide layer around the NW. For
instance, the Zn-Zn distance is 20\% smaller in ZnO than in ZnTe.
The shell thickness is small, \unit{5}{\nano \meter}, resulting in a
small geometrical factor, less than 0.2. Moreover, the shell is
complex, amorphous and probably strongly relaxed. The band edge
emission observed at \unit{2383.7}{\milli \electronvolt} can be
interpreted as the influence of a small residual compressive strain
$f=-0.4 \%$ on the light hole exciton. Hence, even if the nature of
the lines observed in PL here and in Ref.~\onlinecite{Wojn12} cannot
be assessed without a complementary study such as PL excitation or
photoconductivity, their position well agrees with the effect
expected from the strain induced by the shells.

In Fig.~\ref{fig:fig3}.c, we plot the variation of the intensity of
the band-edge emission peak when rotating the linear polarization of
the laser excitation (solid symbols) or of the detection (open
symbols) with respect to the NW axis (determined by SEM, using
substrate marks). The maximum of the emission intensity is reached
when the polarization is parallel to the NW axis (zero angle).
Polarization rates of about 70\% are observed for emission and
detection, in agreement with the value reported with III-V standing
NWs. \cite{Zwil09} These effects result from the dielectric
screening induced by the characteristic aspect ratio of the NWs.
\cite{Wang01}

For small objects such as the present NWs, the CL excitation
efficiency is more than three orders of magnitude lower than the
\textmu PL one. The ratio between the electron-hole excitation
density in CL experiments and in \textmu PL is given by
$\frac{P_{e}/3}{P_{ph}} \times \frac{\eta_{e}}{\eta_{ph}}$. Here
$P_{e}\simeq 2 \mu $W/$\mu $m$^2$ is the mean value of the
electrical power density of the CL electron beam when recording a
typical image, from which is it generally considered that about one
third gives rise to luminescence.\cite{Klein68} $P_{ph} \simeq 25
\mu $W/$\mu $m$^1$ is the laser power density used in the \textmu PL
experiment. The excitation in \textmu PL is non-resonant, 160 meV
above the ZnTe gap, so that we consider that a good order of
magnitude of the absorption by an object of thickness $D$ (the NW
diameter) is given by $\eta_{ph}=\alpha_{ph}D$, where
${\alpha_{ph}\approx4\times10^4}$~cm$^{-1}$ is the value for bulk
ZnTe.\cite{Langen90} The effect of electrons can be calculated using
a Monte-Carlo simulation software (CASINO \cite{CASINO}). An
incident electron has a very small probability to be scattered
inelastically and create an electron-hole pair in the NW; most of
its energy is deposited in the substrate where it is scattered many
times and propagates randomly. As a result, the creation of
electron-hole pairs in the NW is also proportional to \emph{D}, and
much smaller, with an ${\alpha_e\approx4\times10^2}$~cm$^{-2}$. This
low excitation density prevents any spectral analysis of the CL
images for the small NWs studied in this work. Note finally that the
CL signal integrated over the NW width corresponds to an energy
deposited in the NW proportional to ${D^2}$.

\begin{figure}
 \centering
 \includegraphics[width=\columnwidth]{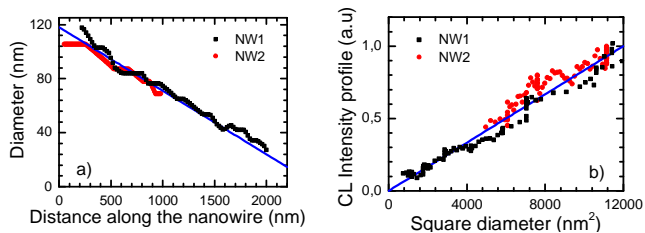}
\caption{(a) (color online) NW diameter as a function of the
position along the NW axis for the two NWs shown in Fig.~3 and
Fig.~4.  (b) CL intensity (integrated over a 12 nm $\times$ 120 nm
window) as a function of the square of the NW
diameter.}\label{fig:fig5}
\end{figure}

Low temperature, polychromatic CL images were recorded in order to
assess the homogeneity of the optical properties. CL emission was
systematically observed from the NWs, as shown in
Fig.~\ref{fig:fig4}.b for the three isolated, broken NWs of
Fig.~\ref{fig:fig4}.a. For thin cone-shaped NWs like the one
labelled $NW_1$ in Fig.~\ref{fig:fig3}.a, and $NW_2$ in
Fig~\ref{fig:fig4}, the CL intensity decreases from the base to the
tip. The local diameter $D$ of $NW_1$ and $NW_2$, determined from
the high resolution SEM images, are shown in Fig.~\ref{fig:fig5}.a
as a function of the position along the NW. Both NWs display a
similar cone shape with the same angle value. Fig.~\ref{fig:fig5}.b
shows that the CL intensity integrated across the NW diameter at
different positions along its axis is proportional to the square of
the diameter. This suggests that the radiative efficiency remains
constant along the NW, with a CL intensity determined by the
excitation density which is proportional to $D^2$.

To conclude, ultra-low density ZnTe NWs have been grown by MBE at
low temperature. These NWs feature a high crystalline quality that
is reflected in their optical response. CL of single ZnTe NWs
deposited on Si substrates was obtained indicating an homogeneous
radiative efficiency along the NWs. Near band edge PL emission was
observed from single NWs. A small blue shift (2-3 meV) of the PL
energy is attributed to the strain induced by the amorphous oxide
layer surrounding the NWs. This assumption is supported by a
quantitative calculation of the excitonic shifts induced by strain
in core-shell NWs.

Most of this work has been done within the CEA-CNRS joint team
``Nanophysique \& Semiconducteurs". We acknowledge the help of the
technical teams of Institut N\'eel, \emph{Nanofab} (clean room) and
\emph{Optical Engineering} (SEM, S. Pairis), and of J.-P. Poizat.
This work is supported by the French National Agency (Magwires,
ANR-11-BS10-013).


\begin{thebibliography}{0}%
\makeatletter
\providecommand \@ifxundefined [1]{%
 \@ifx{#1\undefined}
}%
\providecommand \@ifnum [1]{%
 \ifnum #1\expandafter \@firstoftwo
 \else \expandafter \@secondoftwo
 \fi
}%
\providecommand \@ifx [1]{%
 \ifx #1\expandafter \@firstoftwo
 \else \expandafter \@secondoftwo
 \fi
}%
\providecommand \natexlab [1]{#1}%
\providecommand \enquote  [1]{``#1''}%
\providecommand \bibnamefont  [1]{#1}%
\providecommand \bibfnamefont [1]{#1}%
\providecommand \citenamefont [1]{#1}%
\providecommand \href@noop [0]{\@secondoftwo}%
\providecommand \href [0]{\begingroup \@sanitize@url \@href}%
\providecommand \@href[1]{\@@startlink{#1}\@@href}%
\providecommand \@@href[1]{\endgroup#1\@@endlink}%
\providecommand \@sanitize@url [0]{\catcode `\\12\catcode `\$12\catcode
  `\&12\catcode `\#12\catcode `\^12\catcode `\_12\catcode `\%12\relax}%
\providecommand \@@startlink[1]{}%
\providecommand \@@endlink[0]{}%
\providecommand \url  [0]{\begingroup\@sanitize@url \@url }%
\providecommand \@url [1]{\endgroup\@href {#1}{\urlprefix }}%
\providecommand \urlprefix  [0]{URL }%
\providecommand \Eprint [0]{\href }%
\providecommand \doibase [0]{http://dx.doi.org/}%
\providecommand \selectlanguage [0]{\@gobble}%
\providecommand \bibinfo  [0]{\@secondoftwo}%
\providecommand \bibfield  [0]{\@secondoftwo}%
\providecommand \translation [1]{[#1]}%
\providecommand \BibitemOpen [0]{}%
\providecommand \bibitemStop [0]{}%
\providecommand \bibitemNoStop [0]{.\EOS\space}%
\providecommand \EOS [0]{\spacefactor3000\relax}%
\providecommand \BibitemShut  [1]{\csname bibitem#1\endcsname}%
\let\auto@bib@innerbib\@empty
\end{thebibliography}%


\begin{references}

\bibitem{Duan01} X. F. Duan, Y. Huang, Y. Cui, J. F. Wang and C. M. Lieber, Nature \textbf{409},
66 (2001).

\bibitem{Fan09} Z. Fan, D. J. Ruebusch, A. A. Rathore, R. Kapadia, O. Ergen, P. W.
Leu, and A. Javey, Nano. Res. \textbf{2}, 829 (2009).

\bibitem{Meng09} Q. F. Meng, C. B. Jiang and S. X. Mao, Appl. Phys. Lett., \textbf{94}, 043111 (2009).

\bibitem{Boun12} S. Bounouar, M. Elouneg-Jamroz, M. den Hertog, C. Morchutt, E. Bellet-Amalric, R. Andre
C. Bougerol, Y. Genuist, J.-Ph. Poizat, S. Tatarenko, and K.
Kheng, Nano. Lett. \textbf{12}, 2977 (2012).


\bibitem{Liu13} Z. Liu, G. Chen, B. Liang, G. Yu, H. Huang, D. Chen and G. Shen, Opt. Express \textbf{21}, 7799 (2013).

\bibitem{Rada05} P. V. Radovanovic, C. J. Barrelet, S. Gradecak, F. Qian and C.
M. Lieber Nano. Lett. \textbf{5}, 1407 (2005).

\bibitem{Wojn12} P. Wojnar, E. Janik, L. T. Baczewski, S. Kret, E. Dynowska, T.
Wojciechowski, J. Suffczynski, J. Papierska, P. Kossacki, G.
Karczewski, J. Kossut, and T. Wojtowicz, Nano. Lett. \textbf{12},
3404-3409, (2012).

\bibitem{Bouravleuv13} A. Bouravleuv, G. Cirlin, V. Sapega, P. Werner, and A. Savin, J.
Appl. Phys. \textbf{113}, 144303 (2013).

\bibitem{Ferrand01} D. Ferrand, J. Cibert, A. Wasiela, C. Bourgognon, S. Tatarenko, G. Fishman, T. Andrearczyk, J. Jaroszynski, S. Kolesnik, T. Dietl, B. Barbara, and D. Dufeu
Phys. Rev. B \textbf{63}, 85201 (2001).

\bibitem{Wojn11} P. Wojnar,
E. Janik, L. T. Baczewski, S. Kret, G. Karczewski, T. Wojtowicz, M.
Goryca, T. Kazimierczuk, and P. Kossacki, Appl. Phys. Lett.
\textbf{99}, 113109 (2011).

\bibitem{Beso04} L. Besombes, Y. Leger, L. Maingault, D. Ferrand, H. Mariette and J. Cibert, Phys. Rev. Lett. \textbf{93}, 207403
(2004).



\bibitem{Rueda13} P. Rueda-Fonseca, E. Bellet-Amalric, P. Stepanov, Y. Genuist, M. Den Hertog, D. Ferrand, K. Kheng, R. Andr\'e, J. Cibert, and S. Tatarenko, EuroMBE workshop, Levi, Finland (2013)

\bibitem{Ebina1978_ZnTeOxide} A. Ebina, K.Asano, and T. Takahashi, Phys. Rev. B  \textbf{18},
4341 (1978).

\bibitem{Kirmse08}    H. Kirmse, W. Neumann, S. Kret, P. Dluzewski, E. Janik, G. Karczewski, and T. Wojtowicz, Phys. Stat. Sol. (c) \textbf{5}, 3780 (2008).


\bibitem{Jani07} E. Janik, P. Dluzewski, S. Kret, A. Presz, H. Kirmse, W. Neumann, W
Zaleszczyk, L.T Baczewski, A. Petroutchik, E. Dynowska, J. Sadowski,
W. Caliebe, G. Karczewski, and T. Wojtowicz, Nanotechnology
\textbf{18}, 475606 (2007).

\bibitem{Dona10} F. Donatini and Le Si Dang, Nanotechnology \textbf{21}, 375303 (2010).

\bibitem{Klein68}   C. A. Klein, J. Appl. Phys. \textbf{39}, 2029 (1968).

\bibitem{Langen90} B. Langen, H. Leiderer, W. Limmer, W. Gebhardt, M. Ruff, and U.
Roessler J. Cryst. Growth \textbf{101}, 718 (1990).

\bibitem{CASINO} http://www.gel.usherbrooke.ca/casino/What.html


\bibitem{LeSi89} Le Si Dang, J. Cibert, Y. Gobil, K. Saminadayar, and S. Tatarenko, Appl. Phys. Lett. \textbf{55}, 235
(1989).

\bibitem{Cama02}J. Camacho, A. Cantarero, I. Hernndez-Caldern and L. Gonzlez, J. of Appl. Phys. \textbf{92}, 6014
(2002).

\bibitem{Aifa07} K. E. Aifantis, A. L. Kolesnikova, and E. Romanov, Phil. Magazine \textbf{87}, 4731
(2007).

\bibitem{Cibe13} D. Ferrand and J. Cibert, unpublished.


\bibitem{Berlincourt} D.~Berlincourt, H.~Jaffe, and L.~R.~Shiozawa, Phys. Rev. \textbf{129}, 1009 (1963).


\bibitem{Hartmann} J.~M.~Hartmann, J.~Cibert, F.~Kany, H.~Mariette, M.~Charleux, P.~Alleyson,
R.~Langer, and G.~Feuillet J. Appl. Phys. \textbf{80}, 6257 (1996).


\bibitem{Zwil09} M. H. M. van Weert, N. Akopian, F. Kelkensberg, U.
Perinetti,
M. P. van Kouwen, J. Gomez Rivas, M. T. Borgstrom, R. E. Algra, M.
A. Verheijen, E. P. A. M. Bakkers, L. P. Kouwenhoven, and V.
Zwiller, Small, \textbf{5}, 21342138, (2009).

\bibitem{Wang01} J. Wang, M.S. Gudiksen, X. Duan, Y. Cui, and C.M. Lieber, Science \textbf{293}, 1455 (2001).

\end{references}

\end{document}